\documentclass{webofc}
\usepackage[varg]{txfonts}   
\begin{document}
\title{Fast Interaction Trigger for the upgrade of the ALICE experiment at CERN: design and performance}
%

\author{\firstname{Alla} \lastname{Maevskaya} for the ALICE collaboration\inst{1}\fnsep\thanks{\email{alla@inr.ru}}}

\institute{Institute for Nuclear Research INR RAS, 117312, Moscow, Russia }

\abstract{%
ALICE (A Large Ion Collider Experiment)  at the CERN LHC is designed to study the properties of the Quark--Gluon Plasma  (QGP) in heavy-ion collisions. In 2019--2020 the upgrade of the LHC will increase the luminosity and the collision rate beyond the design parameters of the current ALICE setup. To be able to benefit from the improved performance of the LHC,  ALICE will upgrade several of its key detector systems including the Fast Interaction Trigger (FIT) . FIT is designed to provide the functionality of the existing forward detectors while retaining or even improving their performance. It will provide minimum bias (MB) trigger with an efficiency higher than 90\% for pp collisions, measure the luminosity for pp and Pb--Pb collisions, and sustain interaction rates up to $1~$MHz and $50~$kHz, respectively. FIT will determine the collision time with a resolution better than $50~$ps  and will be used to measure the event multiplicity, the centrality, and the reaction plane. The detector consists of two arrays of Cherenkov radiators with micro-channel plate photo-multiplier (MCP-PMT) sensors, placed on both sides of the interaction point and of a single large-diameter scintillator ring. This article discusses the main design concepts, detector construction, beam test results, Monte Carlo simulations, and the results of detector performance studies. 
}
\maketitle
\section{Introduction}
\label{intro}
The ALICE experiment  \cite{Ref1} at the CERN LHC is dedicated to the study of the Quark--Gluon Plasma – a hot and dense matter produced in ultra-relativistic heavy-ion collisions. After the LHC upgrade in 2019--2020, the interaction rate will reach $50~$kHz for Pb--Pb and up to $2~$MHz for pp collisions. The physics program during  LHC  Run 3 and 4 will take advantage of the unique properties of the upgraded ALICE detector and concentrate on heavy-flavor mesons and baryons,  quarkonium states,  low-mass dileptons up to very low $p_{\mathrm T}$, jets,  and other rare processes. ALICE will be able to accumulate 10~nb$^{-1}$ of Pb--Pb collisions, 6~nb$^{-1}$ of pp as well as 50~nb$^{-1}$ of p--Pb collisions  \cite{Ref2}. Data will be taken either with a Minimum Bias (MB) trigger or in a continuous mode. All events will be transmitted to the online systems. A new Fast Interaction Trigger detector (FIT) \cite{Ref3}  will be installed.
\begin{figure}[hbtp]
\centering
\includegraphics[width=10cm]{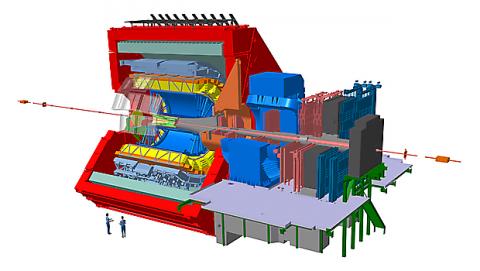}
\caption{Schematic view of the upgraded ALICE detector.}
\label{fig-1}       
\end{figure}

The main components of the upgraded ALICE detector are shown in Fig.~\ref{fig-1}. FIT is required to generate trigger signals with a latency below $425~$ns, provide online luminosity measurements with a direct feedback to the LHC, reject beam-gas background events and determine forward multiplicity of charged particles. During the offline analysis, FIT data will provide centrality, event plane, and the collision time for Time-Of-Flight (TOF) based particle identification. The required time resolution is $50~$ps or smaller for low multiplicity events.

\section{The Fast Interaction Trigger detector}
\label{fit}
\begin{figure}[hbtp]
\centering
\includegraphics[scale=0.25]{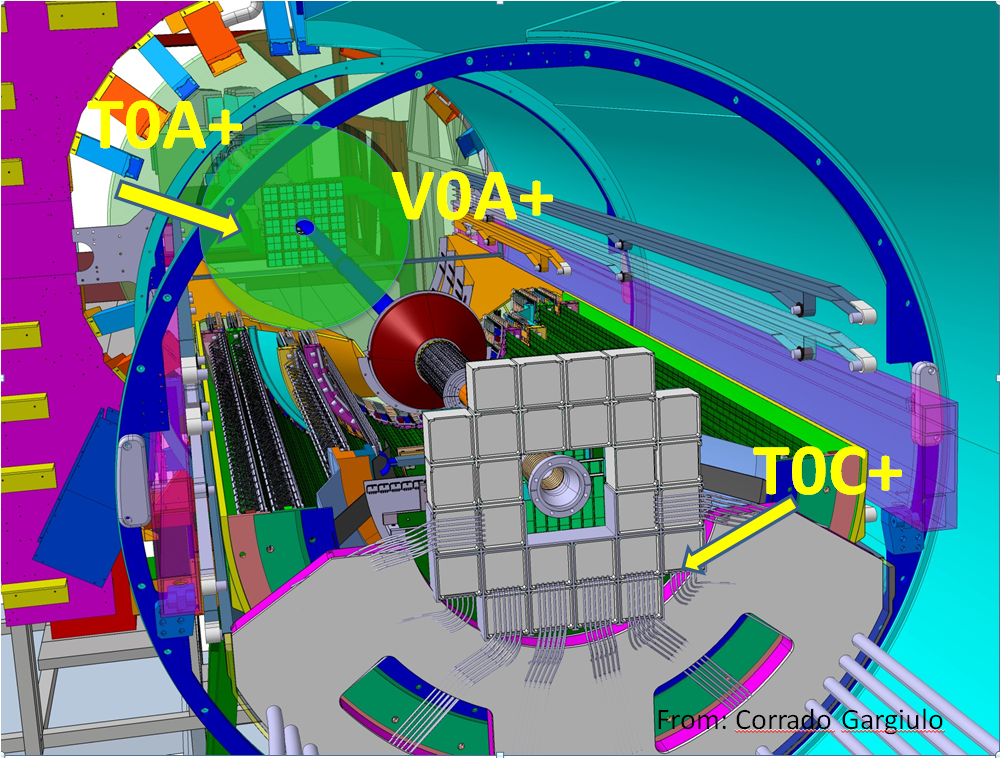}
\includegraphics[scale=0.21]{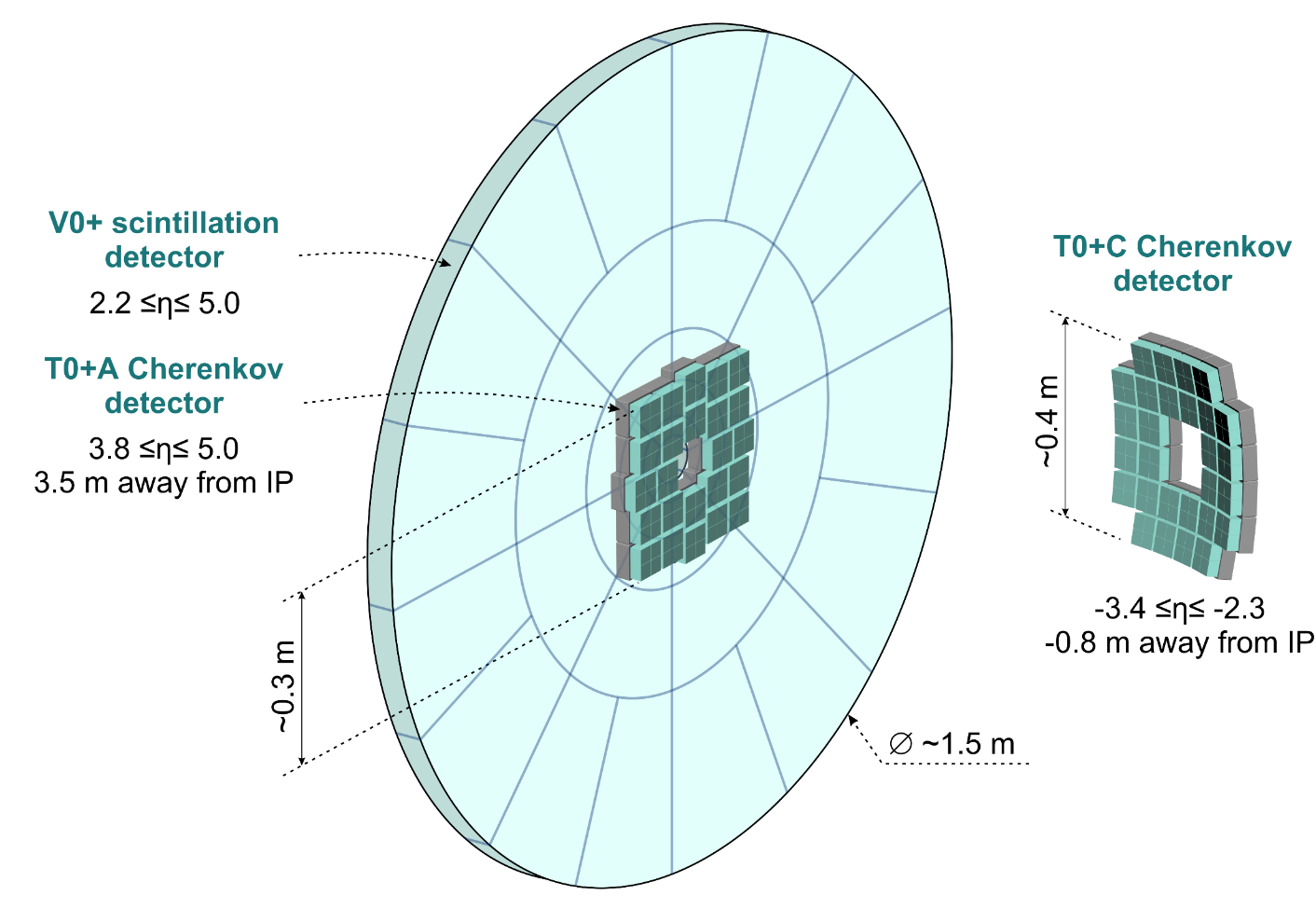}
\caption{Left: View of FIT detector integrated in ALICE. Right: Active components of the FIT detector; the size and pseudorapidity coverage are indicated on the drawing.}
\label{fig-2}       
\end{figure}

To satisfy the requirement of large acceptance and excellent timing, FIT will have a hybrid design. It will consist of two arrays of Cherenkov quartz radiators coupled to fast MCP-PMT photo sensors (T0A+ and T0C+) and of a large, segmented scintillator ring (V0+), as shown on Fig.~\ref{fig-2}. The Cherenkov arrays will be placed  around the beam pipe at 320 cm and –82 cm at the opposite sides of the interaction point (IP) . Because of its proximity to the IP, to equalize the flight time and the effective quartz thickness, T0C+ will have a concave shape with a radius of 82 cm matching the distance to interaction point (IP). The T0A+ and T0C+ arrays will have 24 and 28 modules, respectively. Each T0+ module will consist of 2 cm thick quartz radiators coupled to a modified Planacon XP85012 MCP-PMT. The choice of Planacon XP85012 was based on its performance, compact size (28 mm thick) and a large surface of the photocathode (53 x 53~mm$^{2}$) with respect to the overall face area (59 x 59~mm$^{2}$) of the unit. The XP85012 MCP-PMT has a multi-anode structure (8 x 8 anodes array), which enables segmentation of up to 64 independent pixels. To match the division of the radiators into four  equal elements, the MCP pixels are merged into four  groups of 16 pads (4 x 4) each. In this way each T0+ module operates as four independent detector channels. 
The V0+ detector forms a flat disc with an outer diameter of 148 cm and an inner diameter of 8 cm – just sufficient to pass the beam line through. The active part of V0+ is made of a 4 cm thick EJ-204 plastic scintillators coupled to a grid of clear Asahi fibers transmitting the light to 2” Hamamatsu R5924-70 fine-mesh PMTs.  The scintillator disc is divided into five rings and  eight 45-degree sectors yielding a total of 40 detection elements.

\subsection{FIT electronics}
\label{electronics}

FIT trigger and readout electronics are being developed as a fully integrated system, based on an amplifier, a Constant Fraction Discriminator (CFD), on-board Time and Amplitude to Digital converters (TDC/ADC), FPGA processors, and GBT-based readout. The fact that the trigger decisions will be based on the digitized data (after TDCs \& ADCs) will allow for additional flexibility during the commissioning and operation of FIT.
\subsection{Test results}
\label{test}
T0+ prototypes were tested with a beam of pions and muons with momentum $p = 6~$GeV/${c}$ at the CERN PS. A time resolution of around 33~ps was extracted including the signal deterioration along the 40 m of cables and the actual front-end analog electronics. Tests with a picosecond laser confirmed a linear response of the sensors within the required dynamic range of 1 to 500 particles per quadrant.  According to simulations, the expected average particle load for T0A+ modules closest to the beam pipe in central Pb--Pb collisions will be 280 but the distribution tails may reach up to 500 particles per quadrant.
\subsection{Aging tests}
\label{aging}
During the planned six  years of service the integrated anode current (IAC) accumulated by FIT T0A+ modules closest to the beam pipe will be  $\ge$ 0.57~C/cm$^{2}$. To test the performance of the photosensor a dedicated setup was used. Two quadrants of a Planacon XP85012 were illuminated in sub-saturated mode, while other two quadrants were shaded. After reaching an IAC of 0.5~C/cm$^{2}$, the illuminated quadrant anodes lost ~27\% of the pulse amplitude.  Such a relatively small drop is not problematic as it could easily be compensated by an increase in the high voltage.

For a test in realistic environment a prototype of the FIT T0+ module was installed in the actual ALICE setup and operated there since the beginning of 2016. It yields a very stable time resolution and amplitude, and shows no sign of aging after 2.5 years of operation. 

\subsection{FIT performance studies}
\label{preformance}

\begin{figure}[hbtp]
\begin{center}
\includegraphics[width=12cm]{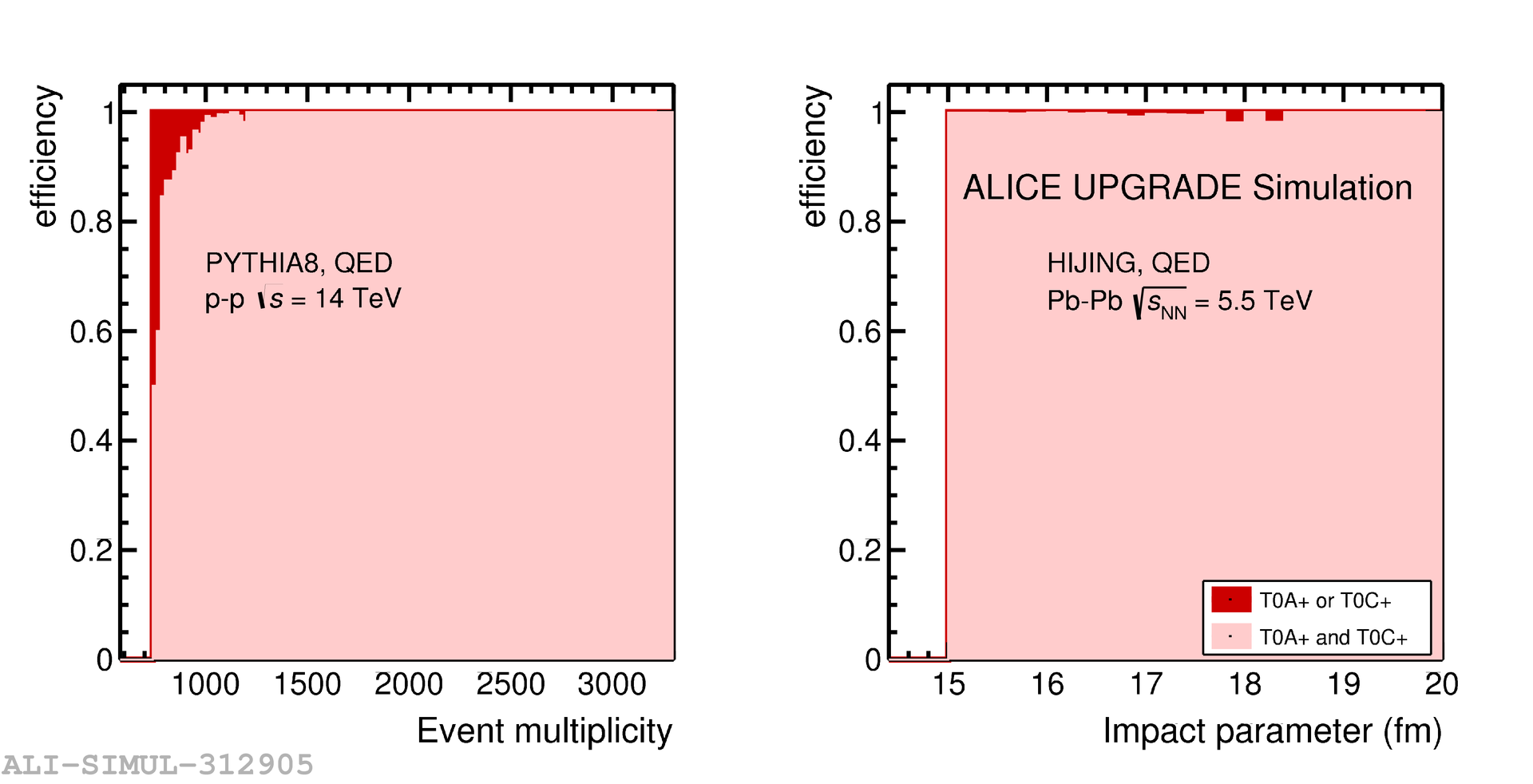}
\end{center}
\caption{MB trigger efficiency  measured by the FIT detector for pp (left) and Pb--Pb (right) collisions}
\label{fig-3}       
\end{figure}

The main requirements concerning FIT are to reach a trigger efficiency of 90\% or higher, and to the maintain the centrality, event plane, and collision time resolution similar to  those achieved during LHC Run 1 and 2. To perform simulation studies the detector geometry and performance parameters obtained during the prototype tests were included into the AliRoot \cite{Ref4} –  ALICE framework for simulation, reconstruction, and data analysis.  AliRoot takes into account all detectors, their support structures, magnetic field, and employs event generators to simulate collision systems.   For pp interactions the Pythia 8 generator  \cite{Ref5} was used and  for Pb--Pb collisions HIJING \cite{Ref6}. Electrons from electromagnetic interactions were added as an additional background. The left panel of Fig.~\ref{fig-3} shows the registration efficiency for pp collisions  as a function of event multiplicity. The right panel shows the efficiency as a function of impact parameter for very peripheral Pb--Pb collisions. The MB trigger conditions used for these calculations were generated as a coincidence between the A and C detectors and OR coincidence. The obtained efficiency was 98.8\% for pp and 99.8\% for peripheral Pb--Pb collisions.

\begin{figure}[hbtp]
\begin{center}
\includegraphics[scale=0.2]{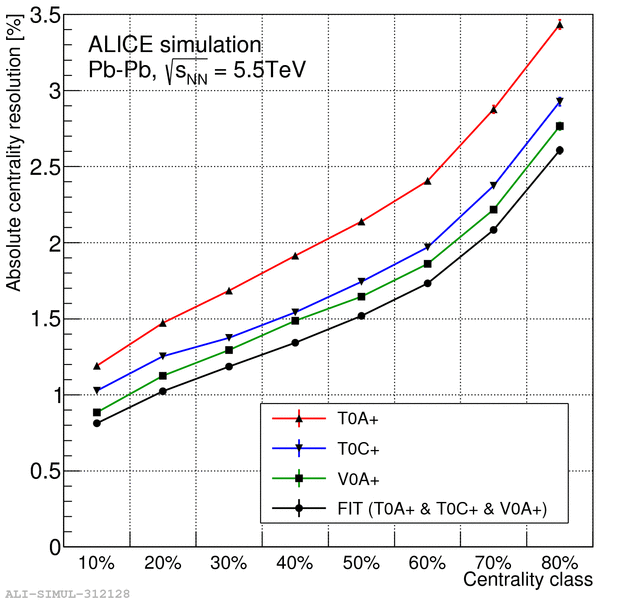}
\includegraphics[scale=0.245]{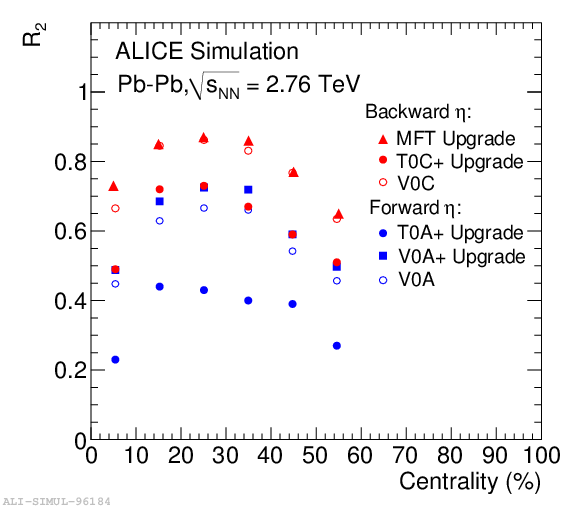}
\end{center}
\caption{Left: Centrality resolution of FIT in Run 3, Pb--Pb collisions at $\sqrt{\it{s}_{\rm{NN}}} = 5.5~$TeV. Right: FIT and MFT (upgrade) performance on event plane resolution R2. }
\label{fig-4}     
\end{figure}

To extract the centrality resolution 50000 Pb--Pb collisions within a centrality range 0--90\% at $\sqrt{\it{s}_{\rm{NN}}} = 5.5~$TeV were generated with HIJING.  The left panel of Fig.~\ref{fig-4} shows the centrality resolution for each component of FIT – T0A+, T0C+ and V0A+ and their combination. One could see that the centrality resolution obtained with the whole detector is below 1\% for central events and increases  to 2.5\% for the most peripheral collisions.
The event plane was estimated following the method of  \cite{Ref7}. The generator included as inputs the direct and elliptic flow distributions from the data. The right panel of Fig.~\ref{fig-4} shows the event plane resolution for the current and upgraded forward detectors. The simulations demonstrate that the resolution obtained with FIT and Muon Forward Tracker (MFT) \cite{Ref8} detectors is equivalent to that of the current detectors.

\begin{figure}
\begin{center}
\includegraphics[width=4cm,height=3cm]{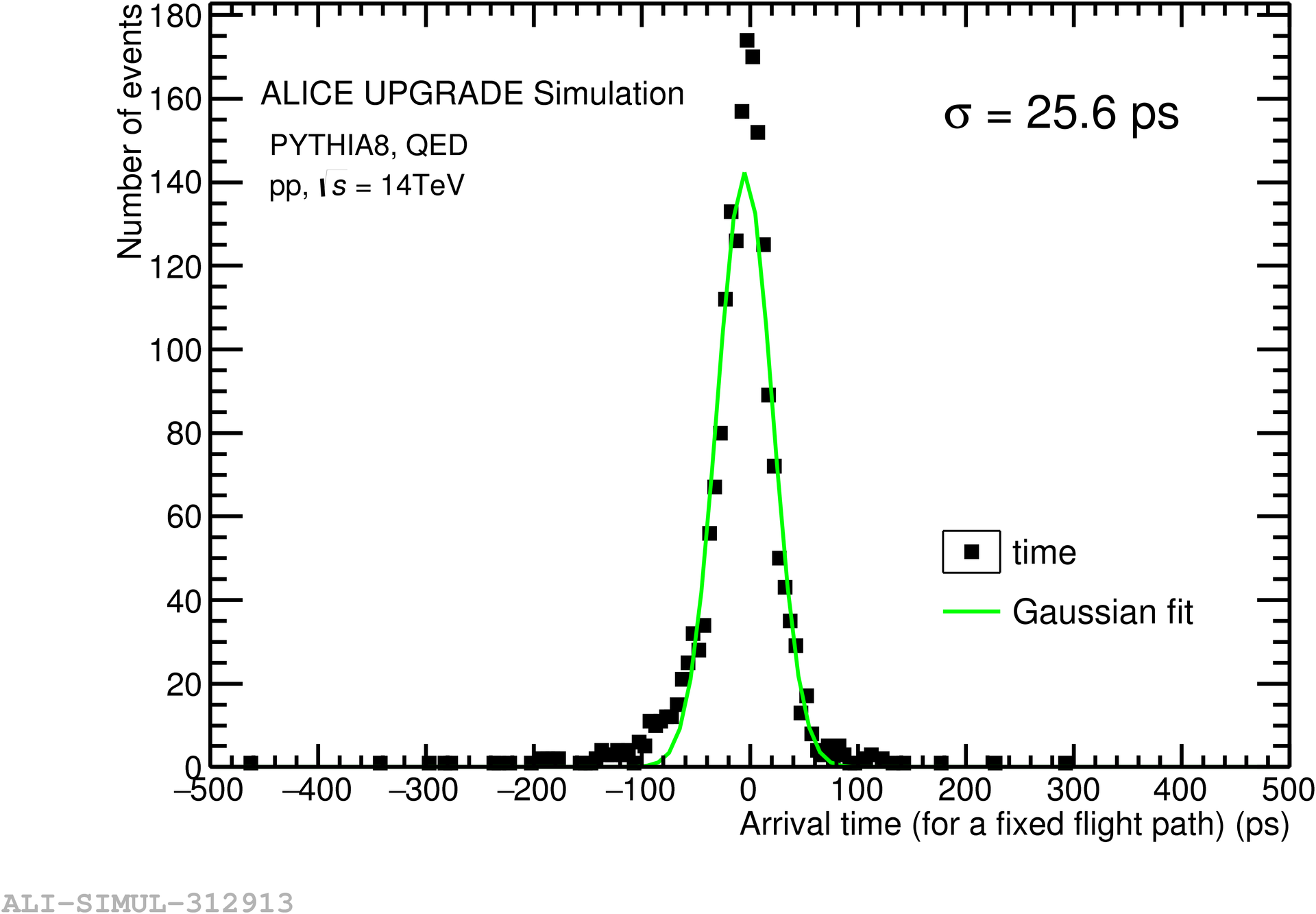}
\includegraphics[width=4.5cm,height=3.33cm]{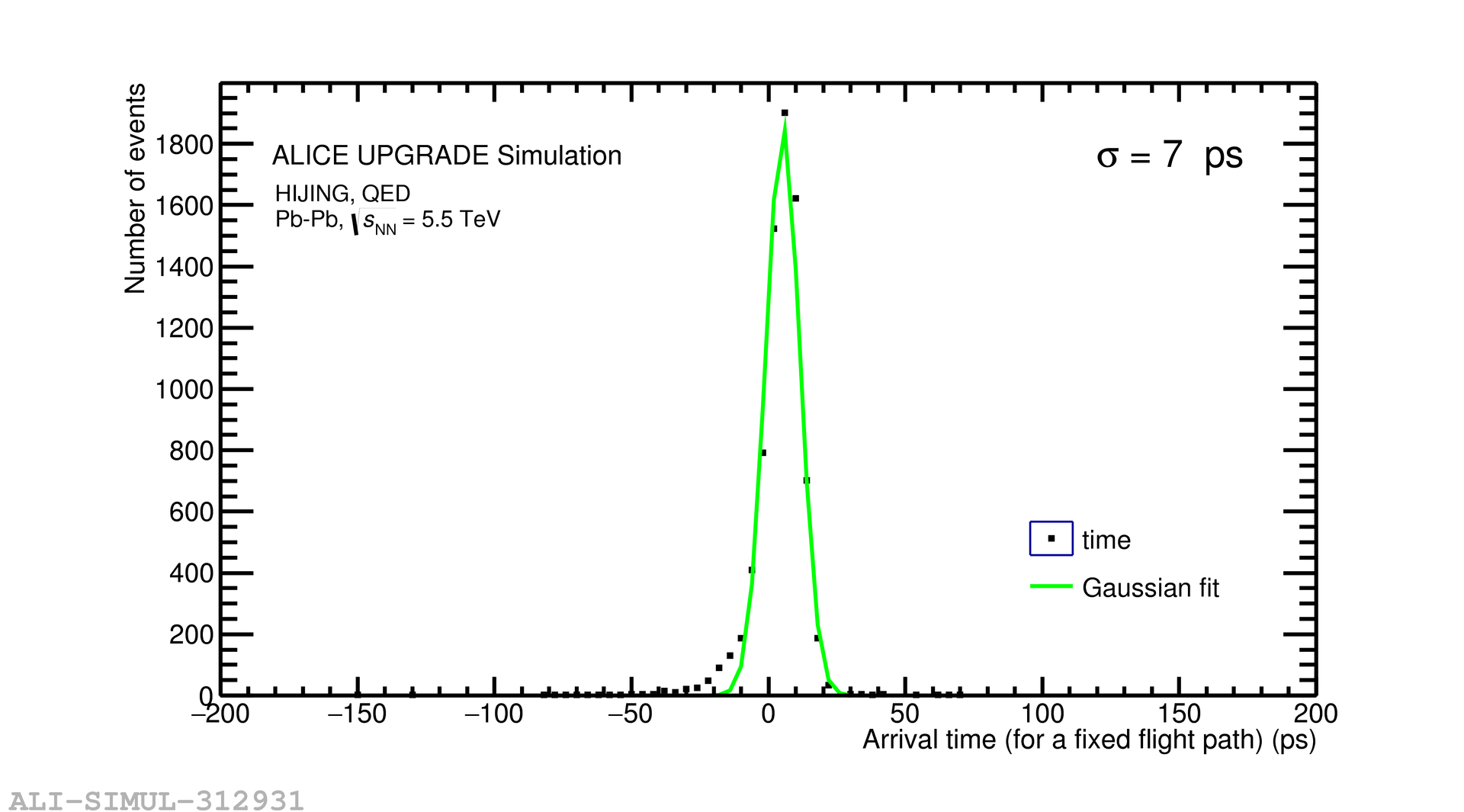}
\end{center}
\caption{Collision time resolution of FIT in Run 3for pp collisions $\sqrt{\it{s}} = 14~$TeV (left) and   Pb--Pb collisions, $\sqrt{\it{s}_{\rm{NN}}} = 5.5~$TeV (right) }.
\label{fig-5}       
\end{figure}

The determination of the collision time with high precision is a
 very important part of the TOF-based particle identification. The collision time
resolution is shown on the left panel of Fig.~~\ref{fig-5} for pp collisions at $\sqrt{\it{s}}$ = 14 TeV and on the right panel  - for  Pb--Pb collisions at  $\sqrt{\it{s}_{\rm{NN}}} = 5.5~$TeV  in the centrality range 0--90\%.

\section{Conclusions}
\label{conc}
The strategy for the upgrade of the ALICE experiment is to benefit fully from the increased LHC luminosity in order to carry out measurements of rare probes requiring triggerless operation. To fulfill that goal several elements of the setup had to be developed including the new Fast Interaction Trigger. FIT prototype tests at the CERN PS have demonstrated that the required functionality can be achieved. The time resolution and reliability of the new detector will be improved retaining the required efficiency, centrality, and event plane resolution. The conducted aging tests of the modified Planacon XP85012  MCP-PMTs predict only a modest ~27\% drop  in the signal amplitude over the period of Run 3.  The sensor fulfills also the linearity requirements over the full amplitude range.

\section{Acknowledgements}
\label{sec-6}
This work was supported for the INR RAS and National Research Nuclear University MEPhI participants within the Program of Russian groups activities in the ALICE upgrade by the Ministry of Education and Science of Russian Federation, contract No14.610.21.0003.

%
%
%

\end{document}